\documentstyle[12pt,group21]{article}

\title{The spectrum of random matrices and integrable
systems
}

\author{Pierre van Moerbeke}
\address{Department of Mathematics, Universit\'e de Louvain,
1348 Louvain-la-Neuve, Belgium 
and Brandeis University, Waltham, Mass 02254, USA
\newline      (vanmoerbeke@geom.ucl.ac.be 
\& vanmoerbeke@math.brandeis.edu).}

\newcommand{\MAT}[1]{\left(\begin{array}{*#1c}}
\newcommand{\mat}{\end{array}\right)}

\newcommand{\rg}{\rightarrow}

\newcommand{\AR}{{\cal A}}
\newcommand{\PR}{{\cal P}}

\newcommand{\BZ}{{\Bbb Z}}
\newcommand{\BC}{{\Bbb C}}

\newcommand{\iy}{\infty}
\newcommand{\pl}{\partial}
\newcommand{\al}{\alpha}
\newcommand{\be}{\beta}
\newcommand{\om}{\omega}

\newcommand{\la}{\langle}
\newcommand{\ra}{\rangle}

\newcommand{\dt}{\delta}

\newcommand{\sg}{\sigma}
\newcommand{\BR}{{\Bbb R}}
\newcommand{\lb}{\lambda}

\def\BX{{\Bbb X}}
\def\BJ{{\Bbb J}}

\def\diag{\mathop{\rm diag}}
\def\Tr{\mathop{\rm Tr}}

\def\span{\mathop{\rm span}}
\def\Im{\mathop{\rm Im}}

\def\be{\begin{equation}}
\def\ee{\end{equation}}
\def\bea{\begin{eqnarray}}
\def\eea{\end{eqnarray}}

\ifx\undefined\Bbb
        \let\Bbb\bf
\fi
\ifx\undefined\curvearrowright
        
\else
        
\fi

\begin{document}
\maketitle

Consider\footnote{Appeared in: Group 21, Physical applications and
Mathematical aspects of Geometry, Groups and Algebras, Vol.II,
835-852, Eds.:H.-D. Doebner, W. Scherer, C. Schulte, World scientific,
Singapore, 1997.} a weight $\rho(dz):=e^{-V(z)}dz$ on an interval 
$F=[A,B]\subseteq\BR$, 
with rational logarithmic derivative and subjected to the
following boundary conditions:
\be
V'=\frac{g}{f}=\frac{\sum_0^{\iy}b_iz^i}{\sum_0^{\iy}a_iz^i},
\quad
\lim_{z\rightarrow
A,B}f(z)e^{-V(z)}z^k=0\mbox{\,\,for all\,\,}k\geq 0.
\ee
On the ensemble
$$
{\cal H}_N=\{N\times N\mbox{\,\,Hermitean matrices}\}
$$
define the probability
\be
P(M\in dM)=c_N e^{-\Tr V(M)}dM,
\ee
with the Haar measure $dM$ on ${\cal H}_N$; the latter can be
decomposed into a spectral part (radial part) and an angular
part:
\be
dM:=\displaystyle{\prod_1^NdM_{ii}\prod_{1\leq i<j\leq
N}}(dReM_{ij}\,d\Im M_{ij})=\Delta^2(z)dz_1...dz_N~dU,
\ee
where $\Delta(z)=\displaystyle{\prod_{1\leq
i<j\leq N}}(z_i-z_j)$ is the Vandermonde determinant. This
lecture deals with the following probability for $E\subset
F$:
\be
P(\mbox{spectrum\,\,}M\in E,\mbox{\,\,with\,\,}M\in{\cal
H}_N)=\frac{\int_{E^N}\Delta^2(z)\prod^N_1\rho(dz_k)}{
\int_{F^N}\Delta^2(z)\prod_1^N
\rho(dz_k)}.
\ee
As the reader can find out from the excellent book by Mehta
\cite{M1}, it is well known that, if the probability
$P(M\in dM)$ satisfies the following two requirements: (i)
invariance under conjugation by unitary transformations
$M\mapsto UMU^{-1}$, (ii) the random variables
$M_{ii}$, Re
$M_{ij}$, Im $M_{ij}$, $1\leq i<j\leq N$ are independent,
then $V(z)$ is quadratic (Gaussian ensemble). For
this ensemble and for very large
$N$, the probability $P$(an eigenvalue $\in dz)$ tends to Wigner's
semi-circle distribution on the interval $[-\sqrt{2N},\sqrt{2N}]$.
Upon normalizing the distribution such that the average
spacing between eigenvalues equals one, one finds different
limits for
$N\nearrow\iy$, according to whether one considers the part of the
spectrum near $z=0$ (bulk scaling limit) or near the edge
$\sqrt{2N}$ (edge scaling limit). The first situation leads to the
Sine kernel and the second one to the Airy kernel.

It is also known that the following probability can be
expressed as a Fredholm determinant, for both, $N$
finite and $N$ infinite. A very sketchy proof of this fact
will be exposed in this lecture. Therefore the real
issue is to compute for
$E\subset F$,
\be
P(\mbox{exactly $k$ eigenvalues in
$E)=\frac{(-1)^k}{k!}\left(\frac{\pl}{\pl\lb}\right)^k\det
(I-\lb K_N^E)\Bigl|_{\lb=1}$}
\ee
for the kernel
\be
K^E(y,z)=K(y,z)I_E(z).
\ee
Below are a few important examples of kernels, the first one being
one side of the Christoffel-Darboux formula for classical orthogonal
polynomials: 
\be
\begin{array}{lll}
\bullet&\displaystyle{
e^{-\frac{1}{2}(V(y)+V(z))}\sum^N_{k=1}p_{k-1}(y)p_{k-1}(z)}&
\mbox{(``Christoffel-}\\
& &\mbox{Darboux" kernel)}\\
\bullet&\displaystyle{\frac{1}{2}\int^x_0(e^{-ixy}\pm
e^{ixy})(e^{ixz}\pm e^{-ixz})dx=\frac{\sin x(y-z)}{y-z}\pm\frac{\sin
x(y+z)}{y+z}}&\mbox{(Sine kernel)}\\
& & \\
\bullet&\displaystyle{\int^x_0A(x+y)A(x+z)dx}&\mbox{(Airy kernel)}\\
& & \\
\bullet&\displaystyle{\frac{1}{2}\int^x_0xJ_{\nu}(x\sqrt{y})
J_{\nu}(x\sqrt{z})dx}&
\mbox{(Bessel kernel).}
\end{array}
\ee
As a feature, common to all of them, note that, for instance,
 the orthogonal polynomials $p_k(y)$, the exponential
function
$e^{-ixy}\pm e^{ixy}$, the Airy function $A(x+y)$ and the Bessel function
$J_{\nu}(x\sqrt{y})$ are all eigenfunctions of second order
problems.

Random matrices provide a model for excitation spectra of
heavy nuclei at high excitations (Wigner\cite{Wig},
Dyson\cite{Dy1,Dy2} and Mehta\cite{M1,M2}). Since the energy
levels are governed by the spectrum of a quantum mechanical
Hamiltonian on a Hilbert space, it is reasonable upon truncation to
assume that the Hamiltonian be represented by large matrices and,
from a statistical mechanical point of view, by an ensemble of
matrices, possibly satisfying some symmetry properties.  In
their analysis of nuclear experimental data, Porter and
Rosenzweig \cite{PR} observed that the occurrence of two levels,
close to each other, is a rare event (level repulsion),
showing that the spacing is not Poissonian, as one might
expect from a naive point of view; this lead Wigner to his
so-called surmise.

In their pioneering work, Jimbo, Miwa, Mori and Sato\cite{JMMS} have
shown some $(p,q)$-variables derived from the sine kernel satisfy a
certain Neumann-like completely integrable finite-dimensional
Hamiltonian system. Tracy and
Widom\cite{TW1,TW2} have successfully used
functional-theoretical tools to compute
the level spacing distributions for a more general class of kernels,
always yielding Neumann-type systems. In the case where
$E$ is a semi-interval, they find a distribution whose density
satisfies a Painlev\'e type equation. Deift, Its and
Zhou\cite{Deift} have used the Riemann-Hilbert approach to find the
precise asymptotics for the distributions above. Random matrices
have come up in the context of statistical mechanics and quantum
gravity; see
\cite{AGL,IIKS,McC}.
 Random matrix ideas play an 
increasingly prominent role in mathematics: 
not only have they come up in the spacings of 
the zeroes of the Riemann zeta function, but their relevance
has been observed in the chaotic Sinai billiard and, more
generally, in chaotic geodesic flows; Sarnak\cite{S}
conjectures that chaos leads to the ``spectral
rigidity", typical of  the spectral distributions of random
matrices,  whereas the spectrum of an integrable system is
random (Poisson)!

The main question of this lecture  is the
following: {\em What is the connection of random matrices with
integrable systems}~? {\em Is this connection really useful}~?
Remember an integrable system is a time evolution involving commuting
vector  fields, parametrized by times
$t_1,t_2,...$\,. Celebrated examples are the KP hierarchy and the
Toda lattice. 

The approach in this lecture, based on joint work with
M. Adler and T. Shiota\cite{ASV2}, is novel and
different from the previous studies, mentioned above. Here we get a
system of partial differential equations directly for the
probability (5)  with $k=0$, rather than a Neumann
system for some complicated auxiliary
variables. Moreover our PDE's are nothing else but the KP hierarchy
for which the
$t$-partials, viewed as commuting operators, are replaced by
non-commuting operators in the endpoints $A_i$ of the interval $E$
under consideration. When the boundary of $E$ consists of one
point, one recovers the Painlev\'e equations, found
in\cite{JMMS,TW1}, at least for some of the kernels (7); from
our work, it also appears that some of the
Painlev\'e equations can be viewed as the KP equation in
non-commutative operators.       

Invoking a different integrable system, H. Peng\cite{P} has
extended these methods to symmetric ensembles, instead of Hermitean
ensembles. Symplectic ensembles will require the use of still a
different integrable system. Picking the 2-Toda lattice leads to
equations for the distribution of the spectrum associated with two
coupled random matrices; see \cite{AvM3}.

The first half of this lecture discusses the finite Hermitean
matrix ensembles, which relate to discrete integrable models, while
the second deals with infinite ensembles, which arise in the
context of continuous systems. The first part serves, to a large
extent, to understand and motivate the methods developped in the
second part.

\section{Finite ensembles at the crossroads of KP, Toda
and Virasoro} 

\subsection{Introducing time in the probability}

On the ensemble ${\cal H}_n$, we consider probability (2), in
which we introduce ``time", seemingly as an artifact:
$$
P(M_n\in dM)=c_n(t)e^{-\Tr V(M)+\sum_1^{\iy}t_i\Tr M^i}dM
$$
and thus
\be
P(\mbox{spectrum\,} M_n\in
E)=\frac{\int_{E^n}\Delta^2(z)\prod^n_{k=1}\rho_t(dz_k)}{
\int_{F^n}\Delta^2(z)\prod^n_{k=1}\rho_t(dz_k)}=:
\frac{\tau_n(E,t)}{\tau_n(t)},
\ee
where
\be
\rho_t(dz)=e^{\sum_1^{\iy}t_iz^i}e^{-V(z)}dz=
e^{\sum_1^{\iy}t_iz^i}\rho_0(dz)
.
\ee

\subsection{$\tau_n(E,t)$ satisfies the KP hierarchy}

Introduce orthonormal and (monic) orthogonal polynomials, 
$p_i(z):=p_i(E,t,z)$ and $\tilde p_i(z):=\tilde p_i(E,t,z)$
respectively, with regard to the density $\rho_t$, on
$E\subset F\subseteq\BR$,
\be
\la p_i,p_j\ra_{E,t}=\delta_{ij},\quad\la\tilde p_i,\tilde
p_j\ra_{E,t}=h_i\delta_{ij},\quad\tilde p_i=\sqrt{h_i}p_i.
\ee
The matrix $m_n$ of moments $\mu_{ij}(E,t)$, at time $t$
\be
m_n(E,t):=\left(\mu_{ij}(E,t)\right)_{0\leq i,j\leq n-1}:=\left(\la
z^i,z^j\ra_{E,t}\right)_{0\leq i,j\leq n-1},
\ee
can be expanded in Schur polynomials $S_k(t)$, with coefficients
given by the moments at time $t=0$:
\be
\begin{array}{ll}
\mu_{ij}(E,t)=\displaystyle{\int_Ez^{i+j}e^{\sum_1^{\iy}t_kz^k}
\rho(dz)}&=\displaystyle{\sum_{k=0}^{\iy}
S_k(t)\int_E z^{i+j+k}}\rho(dz)\\
&=\displaystyle{\sum_{k=0}^{\iy}}S_k(t)\mu_{k+i,j}(E,0).
\end{array}
\ee
Also, it is classically known (see Szeg\" o\cite{Sz}) that both
$\det m_n$ and $p_n(z)$ have integral representations; these
facts  lead to the following identities:
\begin{eqnarray}
\tau_n(E,t)&=&\int_{E^n}\Delta(u)^2\prod^N_{k=1}\rho_t(du_k)
=\det
m_n(E,t)\\
&=&\det\Biggl(\Bigl(S_{j-i}(t)\Bigr)_{_{{0\leq i\leq
n-1}\atop{0\leq j<\iy}}}\Bigl(\mu_{jk}(E,0)\Bigr)_{_{{0\leq
j<\iy}\atop{0\leq k\leq n-1}}}\Biggr)
\end{eqnarray}
and
\begin{eqnarray}
\tilde p_n(E,t,z)&=&\frac{1}{\det
m_n}\int_{E^n}\prod^n_1(z-u_k)\Delta(u)^2\prod^n_{k=1}
\rho_t(du_k)\\
&=&z^n\frac{\tau_n(E,t-[z^{-1}])}{\tau_n(E,t)},
\mbox{\,\,where\,\,}
[\al]=\left(\al,\frac{\al^2}{2},\frac{\al^3}{3},...\right),
\end{eqnarray}
using in the last equation
\be
e^{\sum_1^{\iy}t_iu^i}\left(1-\frac{u}{z}\right)=e^{\sum_1^{\iy}
\left(t_i-
\frac{z^{-i}}{i}\right)u^i}.
\ee
Equation (14) shows that the integral $\tau_n(E,t)$
is the determinant of the projection of an
infinite-dimensional plane onto the positive Fourier modes
$H_+=\span\{1,z,z^2,...\}$, as follows:
\be
e^{\sum_1^{\iy}t_iz^i}\left(\span\left\{\sum^{n-1}_{j=-\iy}z^j
\mu_{n-1-j,k}
(E,0),k=0,...,n-1\right\}\oplus z^nH_+\right)\rg H_+.
\ee
Then, Sato\cite{Date} tells us that this
projection satisfies Hirota bilinear identities (Pl\"ucker
relations) and, in particular, the {\bf
KP-hierarchy}; but according to (13) this projection coincides
with the integral
$\tau_n(E,t)$. Therefore $\tau_n(E,t)$ satisfies for all
$n\geq 0$ the KP-equations\footnote{$S_i(\pm\tilde\pl)=S_i
\left(\pm\frac{\pl}{\pl
t_1},\pm\frac{1}{2}\frac{\pl}{\pl
t_2},\pm\frac{1}{3}\frac{\pl}{\pl t_3},...\right)$, where
$S_i(t)$ are the elementary Schur polynomials:
$\displaystyle{e^{\sum_1^{\iy}t_iz^i}=\sum_0^{\iy}S_k(t)
z^k}.$},
\be
\tau\frac{\pl^2\tau}{\pl t_1\pl t_k}-\frac{\pl\tau}{\pl t_k}
\frac{\pl\tau}{\pl t_1}
-\sum_{{i+j=k+1}\atop{i,j\geq
0}}S_i(\tilde\pl)\tau\cdot S_j(-\tilde\pl)\tau=0,\quad
k=3,4,...\,.
\ee

\subsection{The vector $\left(\tau_n(E,t)\right)_{n\geq 0}$
satisfies the Toda lattice}

Incidentally, equation (16) provides an elementary ``classical"
proof of Sato's representation of a wave-function in terms of a
$\tau$-function. Indeed, in view of equation (23) below, we have that
the $L^2-$norm
$\sqrt{h_k}$  of
$\tilde p$, as in (10), has the form $h_k=\tau_{k+1}/\tau_k$, so that 
$\displaystyle{\tilde
p_k=p_k\sqrt{h_k}=p_k\sqrt{\tau_{k+1}/\tau_k}}$;
this leads to Sato's representation of the so-called ``wave
vector" in terms of the 
``$\tau$-vector"
$\tau(t):=(\tau_n(E,t))_{n \geq 0}$, 
\begin{eqnarray}
\Psi(E,t,z) :=(\Psi_k)_{k\geq
0}&:=&e^{\frac{1}{2}\sum_1^{\iy}t_iz^i}(p_k(E,t,z))_{k\geq
0}\nonumber\\
&=&e^{\frac{1}{2}\sum_1^{\iy}t_iz^i}\left(z^k\frac{\tau_k(t-
[z^{-1}])}{\sqrt{\tau_k(t)\tau_{k+1}(t)}}\right)_{k\geq
0}\nonumber\\
&=&e^{\frac{1}{2}\sum_1^{\iy}t_iz^i}S(t)\chi(z),
\end{eqnarray}
where $\chi(z):=(z^i)_{i\geq 0}$; this last
equation holds, because $p_k(t,z)$ is a triangular linear
combination of monomials $z^j$, represented by the (invertible)
lower-triangular matrix $S(t)$. Note
$z\chi=\Lambda \chi$, where $\Lambda$ is the shift matrix $\Lambda
:=(\dt_{i,j-1})_{i,j\geq 0}$; therefore
\be 
z\Psi=L(t)\Psi,\mbox{\,\,with\,\,}L(t)=S(t)\Lambda
S(t)^{-1}\mbox{\,\,symmetric, tridiagonal matrix.}
\ee
The orthonormality of the polynomials $p_k(E,t,z)$ implies, by (10)
and (20), the orthonormality of the $\Psi_k(E,t,z)$'s for the weight
$\rho_0(dz)$ and thus taking partials in $t_i$, one finds:
$$
\frac{\pl}{\pl
t_i}\int_E\Psi_k(E,t,z)\Psi_{\ell}(E,t,z)\rho_0(dz)=0.
$$
Combining this relation with (20) and (21), one finds that $\Psi$
and
$L$ evolve according to\footnote{with regard to the Lie
algebra splitting of
$g\ell(\iy)$ into the algebras of skew-symmetric $A_s$ and lower
triangular (including the diagonal) matrices $A_b$ 
(Borel matrices):
$$
g\ell(\iy)={\cal D}_s\oplus{\cal D}_b=A=A_s+A_b.$$}:
\be
\frac{\pl\Psi}{\pl t_n}=\frac{1}{2}(L^n)_s\Psi\mbox{\,\,and\,\,}
\frac{\pl L}{\pl
t_n}=\frac{1}{2}[(L^n)_s,L].~\mbox{({\bf Toda lattice})}
\ee
For a full account, see\cite{AvM2}. The seeds for these facts were
already present in the pioneering work of Bessis, Itzykson and
Zuber\cite{BIZ} and later in the work of Witten\cite{Wit}.

\subsection{The $n$-point correlation function}

The methods in this subsection are adapted from Mehta\cite{M1}. 
Return now to probability (8); upon expanding the products of the
two determinants in the integral below, the denominator of (8) can
be expressed as follows, using the orthogonality of the
monic orthogonal polynomials
$\tilde p_k=\tilde p_k(F,t,z)$ on the full range $F$:

\bigbreak

\noindent
$\displaystyle{\int_{F^n}\Delta^2(z)\prod_1^n\rho_t(dz_i)}$
\begin{eqnarray}
&=&\int_{F^n}\det(\tilde
p_{i-1}(z_j))_{1\leq i,j\leq n}\det(\tilde
p_{k-1}(z_{\ell}))_{1\leq k,\ell\leq
n}\prod^n_{n=1}\rho_t(dz_n)\nonumber\\
&=&\sum_{\pi,\pi'\in\sg_n}(-1)^{\pi+\pi'}\prod^n_{k=1}\int_F\tilde
p_{\pi(k)-1}(z_k)\tilde
p_{\pi'(k)-1}(z_k)\rho_t(dz_k)\nonumber\\
&=&n!\prod^{n-1}_{k=0}\int_F\tilde
p^2_k(z)\rho_t(dz)=n!\prod^{n-1}_{k=0}h_k.
\end{eqnarray}

Then,  using the obvious fact $(\det A)^2=\det(AA^{\top})$,
upon borrowing the exponentials from
$\displaystyle{\rho_t(dz_i)}$ and the
$h_i$ from the denominator, one computes the probability (8)
in terms of the polynomials $\tilde p(z):=\tilde
p(F,t,z)$ and a kernel
$K_n(y,z)$:
\bigbreak

\noindent $P(\mbox{spectrum\,} M_n\in E^n)$
\begin{eqnarray}
&=&\frac{1}{n!\prod_1^nh_{i-1}}\int_{E^n}\det\left(\sum_{1
\leq j\leq n}\tilde p_{j-1}(z_k)\tilde
p_{j-1}(z_{\ell})\right)_{1\leq
 k,\ell\leq n}\prod^n_1\rho_t(dz_i)\nonumber\\
&=&\frac{1}{n!}\int_{E^n}\det(K_n(z_k,z_{\ell}))_{1\leq
k,\ell\leq n}
\prod_1^n\rho_0(dz_i),
\end{eqnarray}
where the kernel $K_n$ is defined in terms of the wave function
(20) for $E=F$:
\be
K_n(y,z):=\sum^n_{j=1}\Psi_{j-1}(F,t,y)\Psi_{j-1}(F,t,z).
\ee
The orthogonality relations   
$\displaystyle{\int_F\Psi_k(F,t,z)\Psi_{\ell}(F,t,z)\rho_0(dz)=
\dt_{k\ell}}$ lead to the reproducing property
for the kernel
$K_n(y,z)$:
\be
\int_F K_n(y,z)K_n(z,u)\rho_0(dz)=K_n(y,u),
\quad\quad\int_F K_n(z,z)\rho_0(dz)=n.
\ee
Upon replacing $E^n$ by
$\displaystyle{\prod^k_1}dz_i\times F^{n-k}$ in (24), upon
integrating out all the remaining variables $z_{k+1},...,z_n$ and
using the reproducing property (26), one finds the $n$-point
correlation function 
$$P(\mbox{one eigenvalue in each\,\,}[z_i,z_i+dz_i],i=1,...,k)$$
\be
=c_n\det\left(K_n(z_i,z_j)\right)_{1\leq
i,j\leq k}\prod_1^k\rho_0(dz_i).
\ee
Finally, by Poincar\'e's formula for the probability
$\displaystyle{P\left(\bigcup E_i\right)}$, the
probability that no spectral point of $M$ belongs to $E$ is
given by a Fredholm determinant
$$\det(I-\lb
A)=1+\sum^{\iy}_{k=1}(-\lb)^k\int_{z_1\leq ...\leq
z_k}\det\Bigl(A(z_i,z_j)\Bigr)_{1\leq i,j\leq
k}\prod_1^k\rho_0(dz_i),
$$ for the kernel $A=K_n^E$, given by (25):
\be
P(\mbox{no spectrum\,\,}M\in
E)=\det(I-K_n^E),\quad K_n^E(y,z)=K_n(y,z)I_E(z).
\ee

\subsection{Toda vertex operators and boundary-time Virasoro
constraints}

This subsection is taken from\cite{AvM2,AvM3}; as a new
ingredient, we introduced in\cite{AvM2} the {\it ``vector vertex
operator"} for the Toda lattice,
\be
\BX(t,z):=\Lambda^{-1}\chi(z^2)e^{\sum_1^{\iy}t_iz^i}
e^{-2\sum_1^{\iy}\frac{z^{-i}}{i}\frac{\pl}{\pl t_i}},
\ee
acting on the vector $(\tau_n)_{n\geq 0}$; it can be viewed as a
Darboux transformation, mapping a Toda solution into a new one. It
satisfies the following simple {\it ``operator
identity"}\footnote{$(\BJ_{\ell}\BX\tau)_k=
(\BJ_{\ell})_k(\BX\tau)_k$,
with $(\BX\tau)_k=z^{2(k-1)}e^{\sum_1^{\iy}t_iz^i}
e^{-2\sum_1^{\iy}\frac{z^{-i}}{i}\frac{\pl}{\pl t_i}}
\tau_{k-1}$}:
\be
\left(-b^{\ell+1}\frac{\pl}{\pl b}-a^{\ell+1}\frac{\pl}{\pl
a}+[\BJ_{\ell}^{(2)},\cdot]\right)\int^b_adz\,\BX(t,z)=0,\quad\ell\geq
-1,
\ee
where the vector operators\footnote{the agreement
is the following: $\displaystyle{\frac{\pl}{\pl t_k}=0}$ for $k\leq
0$ and $t_k=0$ for $k\geq 0$. Normal ordering :: means: pull
the differentiation to the right, regardless of commutation
relations.}
\begin{eqnarray}
\BJ_k^{(1)}&=&\frac{\pl}{\pl
t_k}+\frac{1}{2}(-k)t_{-k}+\dt_{k,0}\diag(...,-2,-1,0,1,2,...),\quad
k\in\BZ\nonumber\\
\BJ_k^{(2)}&=&\sum_{i+j=k}\colon\BJ_i^{(1)}\BJ_j^{(2)}\colon,\quad
k\in\BZ
\end{eqnarray}
form the generators of the Heisenberg (oscillator) and Virasoro
algebras respectively.

The Hirota bilinear identities
for $\tau$-functions, already mentioned in section 1.2, also 
generate so-called Fay identities, familiar to algebraic
geometers; they are quadra-\linebreak tic relations between
translates of
$\tau$-functions.  These Fay identities enable us to express the
kernel
$K_n$, defined by (25), and the
$n$-point correlation functions \linebreak $\det(K_n(z_i,z_j))$,  in
terms of vertex operators acting on the $\tau$-vector 
$\tau(t):=(\tau_n(F,t))_{n \geq 0}$, underlying $\Psi(F,t,z)$ as in
(20). Namely,
\be
K_n(z,z)=\sum^n_{j=1}\Psi_{j-1}(t,z)^2=\tau_n^{-1}(\BX(t,z)\tau(t))_n
,\ee 
and, more generally, using higher degree Fay identities:
\be
\det(K_n(z_i,z_j))_{1\leq i,j\leq
k}=\tau_n^{-1}\left(\prod^k_{i=1}\BX(t,z_i)\tau(t)\right)_n.
\ee
Using the Neumann series in $\lambda$ for $\det(I-\lb
K_n^E)$, and setting $E^c:=F\backslash E$, it follows that
\be
P(\mbox{no spectrum\,\,} M_n\in E)=
\frac{\tau_n(E^c,t)}{\tau_n(t)}=
\frac{(e^{-\int_E\BX(t,z)e^{-V(z)}dz}\tau(t))_n}{\tau_n(t)}.
\ee
Given the disjoint union
$$
E=\bigcup^r_{i=1}[A_{2i-1},A_{2i}]\subset F\subset \BR,
$$
the vectors $\tau(t):=\tau(F,t):=(\tau_n(F,t))_{n \geq
0}$ and $\tau(E^c,t):=(\tau_n(E^c,t))_{n \geq 0}$ obey the
following Virasoro-like constraints for
$m=-1,0,1,...$:
\be
\sum_{i\geq
0}\left(a_i\BJ_{i+m}^{(2)}-b_i\BJ^{(1)}_{i+m+1}\right)\tau(F,t)
=0.
\ee
\be
\left(-\sum_1^{2r}A_i^{m+1}f(A_i)\frac{\pl}{\pl A_i}+\sum_{i\geq
0}\left(a_i\BJ_{i+m}^{(2)}-b_i\BJ^{(1)}_{i+m+1}\right)\right)\tau(E^c,t)
=0.
\ee
A sketch of the proof of equations (35) and (36) goes as follows:
besides the matrix
$L$ satisfying $L\Psi=z\Psi$, we introduce a new matrix $M$, a
matrix analogue of the Orlov-Schulman operator for KP, such that
$M\Psi=(\pl/\pl z)\Psi$; of course,
$[L,M]=1$. The boundary condition (1) for the weight $\rho(z) dz$
on $F$ implies the vanishing of the integral
\be
 \int_F \frac{\pl}{\pl z}\left(f(z)\Psi_k(z) \Psi_{\ell}(z)
e^{-V(z)}\right)dz=0.
\ee
Working out this integral and using the operators $L$ and $M$, one
proves the skew-symmetry of the matrix $Q:=Mf(L) + \frac{1}{2}
(f'-g)(L)$, and hence
of\footnote{$[A,B]^{\dagger}=\frac{1}{2}(AB+BA)$}
$[Q,L^{m+1}]^{\dagger}$; remember $f$ and
$g$ are the functions appearing in the weight (1); incidentally, the
skew-symmetric matrix $Q$ actually satisfies the {\em string equation}
$[L,Q]=f(L)$. We then use the ASV-correspondence\cite{AvM2,ASV1}
between symmetry vector fields on the Toda wave functions and the
Virasoro symmetries on the
$\tau$-functions; this establishes equation (35).
Equation (36) follows from (35), using (30) and taking into account
the boundary of $E^c$. Observe the {\em decoupling}  of the
Virasoro constraints in (36) into a {\em time-part} and a {\em
boundary-part}~! 

\subsection{``Non-commutative" KP hierarchy for the
probability distribution}

Let now the rational function $V'(z)$, appearing in the
weight (1), be of the following form:
\be
V'(z)=\frac{g}{f}=\frac{b_0+b_1z}{a_0+a_1z+a_2z^2};
\ee
then the equations (36) and their powers enable us to extract all
partial derivatives
$\pl^{k_1+...+k_{\ell}}\tau(E^c,t)/\pl t_1^{k_1}...\pl
t_{\ell}^{k_{\ell}}$, evaluated at
$t=0$ in terms of partial differential operators in the boundary
points of the disjoint union E, 
\be
{\cal A}_k=\sum_{i=1}^{2r}A_i^{k-m}f(A_i)\frac{\pl}{\pl A_i}\quad
k=1,2,3,...,
\ee
where $m=\max(\deg
f-1,\deg g)$. Declare ${\cal A}_k$ to be of homogeneous weight $k$. 

When $a_2=0$ in $V'$, the probability ${\cal
P}_n(A_1,...,A_{2r}):=P$(no spectrum $M_n\in E)$ satisfies
the KP-hierarchy (19), but with the ({\it commutative})
partial derivatives in $t$ replaced by the ({\it
non-commutative}) partial differential operators ${\cal A}_k$
defined by (39)\footnote{$S_i(\pm\tilde{\cal
A}):=S_i\displaystyle{(\pm{\cal A}_1,\pm\frac{1}{2}{\cal
A}_2,\pm\frac{1}{3}{\cal A}_3,...)}$ are the elementary Schur
polynomials with $t_k$ replaced by
$\displaystyle{\frac{1}{k}}{\cal A}_k$.}:
$$
{\cal P}_n\cdot{\cal A}_k
\AR_1\PR_n-\AR_k\PR_n\cdot\AR_1\PR_n-\sum_{i+j=k+1}S_i(\tilde\AR)
\PR_n\cdot S_j(-\tilde\AR)\PR_n
$$
\be
+ (\mbox{terms of lower weight $i$ for $1\leq i\leq k)=0$,
for $k\geq 3.$}
\ee
When $a_2\neq 0$ in $V'$, the $\AR_k$'s in (36) are replaced
by more complicated expressions.

\medbreak

\noindent\underline{Examples}: When $V(z)=z^2$, the $p_k(\BR,0,z)$'s
are Hermite polynomials and, for a semi-infinite interval
$E=(A,\iy)$, the first equation of the system (40) reduces to the
Painlev\'e IV equation for $\pl /\pl A \log
\det (I-K_n^E)$. The case $V=z-\al \log z$ leads to
Laguerre polynomials at $t=0$ and to Painlev\'e V for
$(A~\pl /\pl A)\log \det (I-K_n^E)$. The Jacobi polynomials are
also a special case of (38) and the first equation of the
hierarchy (40) reduces to an unknown equation. These
Painlev\'e equations were first obtained by Tracy and Widom\cite{TW1}
and then by us\cite{ASV2}, as a special instance of the
non-commutative KP. 

\section{Continuous kernels, KP and Virasoro}

Here, we shall be dealing with continuous kernels, like the
Sine, Airy or Bessel kernels listed in (7). How do we obtain
PDE's for the spectral distribution, using the theory of integrable
systems~? The question is how to introduce time~?  This question
was resolved by Adler, Shiota and the author in\cite{ASV2}. In the
finite situation, the kernel
$K_N$ in (25) was given by a Christoffel-Darboux sum, based on
eigenvectors of
$L$:
\be
K_n(y,z):=\sum^n_{j=1}\Psi_{j-1}(t,y)\Psi_{j-1}(t,z),
\ee
with $\Psi:=(\Psi_j)_{j\geq 0}$ an eigenvector of a second
order problem and evolving in time according to the Toda
lattice (see (21) and (22)):
\be
z\Psi=L\Psi,\quad\frac{\pl\Psi}{\pl
t_n}=\frac{1}{2}(L^n)_s\Psi,\quad\frac{\pl L}{\pl
t_n}=\frac{1}{2}[(L^n)_s,L].
\ee
The {\em continuous analogue} goes as follows: consider now wave
and adjoint wave functions $\Psi(x,t,z)$ and
$\Psi^{*}(x,t,z)$, with $x \in \BR,~t \in \BC^{\iy},~ z\in \BC$
satisfying the KP-hierarchy,
\be
\begin{array}{lll}
z\Psi=L\Psi,&\displaystyle{\frac{\pl\Psi}{\pl
t_n}}=(L^n)_+\Psi, & \\
  &  &  \displaystyle{\frac{\pl L}{\pl
t_n}}=[(L^n)_+,L],\nonumber\\
z\Psi^*=L^{\top}\Psi^*,&\displaystyle{\frac{\pl\Psi^*}{\pl
t_n}}=-(L^{\top n})_+\Psi^*, & 
\end{array}
\ee 
where $L$ is a pseudo-differential operator
$$
L=D+a_{-1}D^{-1}+...,\quad \mbox{with}\quad D=\frac{\pl}{\pl x}.
$$
We consider the $p$-reduced KP hierarchy, i.e., the reduction
to $L$'s such that
$L^p=D^p+\ldots$ is a differential operator for some
$p\geq 2$. The precise continuous analogue of
(41), which relates to a second order problem, would be to choose
$p=2$. 

For the time being, we take $p\geq 2$ arbitrary.
According to Sato's theory, we have the following
representation of $\Psi$ and $\Psi^*$ in terms of a
$\tau$-function (see\cite{Date})
\be
\Psi(x,t,z)=e^{xz+\sum_1^{\iy}t_iz^i}
\frac{\tau(t-[z^{-1}])}{\tau(t)},\quad\mbox{and}\quad
\Psi^*(x,t,z)=e^{-xz-\sum_1^{\iy}t_iz^i}
\frac{\tau(t+[z^{-1}])}{\tau(t)};
\ee
for notation $[z^{-1}]$, see (16). Observe\footnote{$\zeta_p:=\{\om
\mbox{ such that }
\om^p=1\}$}
\be
\Phi(x,t,z):=\sum_{\om\in\zeta_p}b_{\om}\Psi(x,t,\om z)\mbox{ and
}\Phi^*(x,t,z)=\sum_{\om\in\zeta_p}a_{\om}\Psi^*(x,t,\om z),
\ee
are the most general solution of the spectral problems
$L^p\Phi=z^p\Phi$ and $L^{\top p}\Phi^*=z^p\Phi^*$ respectively.

The continuous analogue of (41) is
$$
K_x(y,z):=\int^xdx~\Phi^*(x,t,y)\Phi(x,t,z),\mbox{
also } K^E_x(y,z):=K_x (y,z)I_E(z),
$$
where $\Phi$ and $\Phi^*$ are given by (45), subjected to
the condition $\sum_{\om \in \zeta_p}\frac{a_{\om}b_{\om}}{\om}=0$.
Introduce the p-reduced vertex operator
$$
Y(t,y,z):=\sum_{\om,\om'\in\zeta_p}c_{\om\om'}X(t,\om y,\om'z),
$$
which maps the space of $p$-reduced KP $\tau$-functions into
itself, with
$$
X(t,y,z):=\frac{1}{z-y}e^{\sum_1^{\iy}(z^i-y^i)t_i}
e^{\sum_1^{\iy}(y^{-i}-z^{-i})\frac{1}{i}\frac{\pl}{\pl
t_i}};
$$
see\cite{Date,vM}. We now set
$c_{\om\om'}=a_{\om}b_{\om'}$ in the operator $Y$ above; again
using Fay identities, we obtain continuous analogues of (32),(33)
and (34),
$$K_x(y,z):=\frac{1}{\tau(t)}Y(t,y,z)\tau(t)$$
$$\det\Bigl(K_x(y_i,z_j)\Bigr)_{1\leq i,j\leq k}
=\frac{1}{\tau}\prod^k_{i=1}Y(t,y_i,z_i)\tau$$
$$\det(I-\lb
K_x^E)=\frac{1}{\tau}e^{-\lb\int_Edz\,Y(t,z,z)}\tau.$$
If a $\tau$-function $\tau(t)$ satisfies a Virasoro
constraint\footnote{$J^{(1)}_{\ell} = \frac{\pl}{\pl
t_{\ell}}+(-\ell)t_{-\ell},~J^{(2)}_{\ell}
=\sum_{i+j=\ell}:J^{(1)}_iJ^{(1)}_j:; $ see footnote 4.} (analogue
of (35)):
$$
J_{kp}^{(2)}\tau=c_{kp}\tau\quad\mbox{for a fixed $k\geq -1$}
$$
then, given a subset
$E=\displaystyle{\bigcup^r_{i=1}}[A_{2i-1},A_{2i}]\subset\BR_+$,
the Fredholm determinant of
$$
\tilde K^E(\lb,\lb'):=\frac{1}{p}
\frac{k_{x,t}(z,z')}{z^{\frac{p-1}{2}}z^{'\frac{p-1}{2}}}I_E(\lb'),\quad\lb
=z^p,\lb'=z^{'p}
$$
satisfies the following constraint
$$
\left(-p\sum^{2r}_{i=1}A_i^{k+1}\frac{\pl}{\pl
A_i}+\frac{1}{2}(J_{kp}^{(2)}-c_{kp})\right)\tau\det(I-\mu
\tilde K_x^E)=0.
$$
Notice again the {\em decoupling of the boundary- and the
time-parts}, as in (36). We further proceed in the same
way as in the discrete case: Under suitable assumptions on the
initial conditions, analogous to (38), we can express all
$t$-partials at
$t=0$ in terms of partial differential operators in the endpoints
$A_i$ of $E$, provided $p=2$; we then substitute those expressions
in the KP hierarchy, thus leading to a hierarchy of PDE's in terms
of non-commutative operators in the $A_i$. A precise
statement due to Adler-Shiota-van Moerbeke\cite{ASV2} proceeds as
follows:

\proclaim Theorem.  Consider a first-order differential
operator $A$ in $z$ of the form
\be
A=A_z=\frac{1}{2}z^{-m+1}\left(\frac{\pl}{\pl
z}+V'(z)\right)+\sum_{i\geq 1}c_iz^{-2i},
\ee
with
\be
V(z)=\frac{\al}{2}z+\frac{\beta}{6}z^3\not\equiv 0,\quad
m=\deg V'=0\mbox{\,\,or\,\,}2,
\ee
and its ``Fourier" transform
\be
\hat A=\hat
A_x=\left(\frac{1}{2}\left(x+V'(D)\right)D^{-m+1}+\sum_{i\geq
1}c_iD^{-2i}\right)_+\quad \mbox{with} \quad D=\frac{\pl}{\pl x}.
\ee
Let $\Psi(x,z)$, $x\in\BR$, $z\in\BC$ be a solution
of the linear partial differential equation
\be
A_z\Psi(x,z)=\hat A_x\Psi(x,z),
\ee
with holomorphic (in $z^{-1}$) initial condition at $x=0$, subjected
to the following differential equation for some $a,b,c\in \BC$,
\be
(aA^2+bA+c)\Psi(0,z)=z^2\Psi(0,z),\mbox{\,\,with\,\,}
\Psi(0,z)=1+O(z^{-1}).
\ee
Then\newline\indent $\bullet~~~\Psi(x,z)$ is a
solution of a second order problem for some potential $q(x)$
\be
(D^2+q(x))\Psi(x,z)=z^2\Psi(x,z).
\ee
\indent $\bullet~~~$ Given a subset
$E=\displaystyle{\bigcup^r_{i=1}}[A_{2i-1},A_{2i}]$ and the
kernel
\be
K^E_x(y,z):=I_E(z)\int^x
\frac{\Phi(x,\sqrt{y})\Phi(x,\sqrt{z})}{2y^{1/4}z^{1/4}}
dx,
\ee
with
$$
\Phi(x,u):=\sum_{\om=\pm 1}b_{\om}e^{\om V(u)}\Psi(x,\om
u),
$$
the Fredholm determinant ${\cal P}(A_1,...,A_{2r}):=\det(I-\lb
K_x^E)$ satisfies a hierarchy of bilinear partial
differential equations\footnote{$S_i(\pm\tilde{\cal
A}):=S_i\displaystyle{(\pm{\cal A}_1,0,\pm\frac{1}{3}{\cal
A}_3,0,...)}$} in the $A_i$ for odd $n\geq 3$:
$$
{\cal P}\cdot\AR_n\AR_1{\cal
P}-\AR_n{\cal P}\cdot\AR_1{\cal P}-\sum_{i+j=n+1}S_i
(\tilde\AR){\cal P}\cdot S_j(-\tilde\AR){\cal P}$$
\be
+ \,(\mbox{terms of lower weight $i$ for $1\leq i\leq n)=0$},
\ee

\medbreak

\noindent where the $\AR_n$ are differential operators of
homogeneous ``weight"
$n$, defined by
$$
\AR_n=\sum_{i=1}^{2r}A_i^{\frac{n+1-m}{2}}\frac{\pl}{\pl A_i},
\quad n=1,3,5,... .
$$

\medbreak

\noindent\underline{examples}. {\bf Airy kernel}: Given an entire
function
$U(u)$ growing sufficiently fast at $\iy$, consider its ``Fourier"
transform,
\begin{equation}
F(y)=\int_{-\iy}^\iy e^{-U(u)+uy}du~;
\end{equation} define the associated function
$\rho(z)$ and the differential operator $A$:
\be
\rho(z)=\frac{1}{\sqrt{2\pi}}e^{U(z)-zU'(z)}\sqrt{U''(z)}\mbox{ and
} A:=\rho(z) \frac{1}{U''(z)}\frac{\pl}{\pl z}  \rho(z)^{-1}. 
\ee
Then the function
\be
\Psi(x,z):= \rho(z) F(x+U'(z))
\ee
obeys, for a trivial reason,
\be
A\Psi(x,z)=\frac{\pl}{\pl x}\Psi(x,z),
\ee
and, upon integration of $\frac{\pl}{\pl u}e^{-U(u)+uy}$,
one shows $\Psi(x,z)$ also satisfies:
\be
\left(U'(\frac{\pl}{\pl x})-x\right)\Psi(x,z) = U'(z) \Psi(x,z).
\ee
Upon letting $x\rightarrow 0$ in (58) and using (57), we find
\be
U'(A)\Psi(0,z)=U'(z)\Psi(0,z).
\ee
Choosing $U(z)=z^3/3$, the integral (54) becomes the Airy
function; the operator $A$ in (55) coincides with (46),
for $V(z)=\frac{2}{3}z^3, ~\al=0,~\beta=4,~m=2$: 
$$
A:=\frac{1}{2z}\left( \frac{\pl}{\pl
z}+2z^2\right)-\frac{1}{4}z^{-2}
~\mbox{and}~ \hat A=D.
$$
Moreover (59) becomes equation (50) and, by stationary phase,
satisfies the asymptotics requested in (50):
$$A^2 \Psi(0,z)=z^2\Psi(0,z),~\mbox{with}~
\Psi(0,z)=1+O(z^{-1})~~\mbox{for}~~z
\nearrow \iy.
$$
From (58), it follows that $q(x)=-x$ in (51); the kernel (52)
reduces to the Airy kernel mentioned in (7) and therefore its
Fredholm determinant satisfies hierarchy (53) with ${\cal
A}_n=\sum_{i=1}^{2r}A_i^{\frac{n-1}{2}} \pl/\pl A_i$.  When $E$ is
a semi-infinite interval, the first PDE in the hierarchy (53)
reduces to Painlev\'e II for $(\pl/\pl A) \log \det (I-\tilde
K^E)$, as originally discovered in\cite{TW1}.  Note that in this
example, the
$\tau$-function is the Kontsevich integral.  For details on our
methods, see\cite{AMSV,ASV2}.

\medbreak

\noindent {\bf Bessel kernel}: Pick $m, V(z), A_z$ and $\hat{A}_x$ in
(46), (47) and (48) as follows,
$$
V(z)=-z, \mbox{ with } m=0, A:=A_z:=\frac{1}{2} z(\frac{\pl}{\pl
z}-1), \hat{A}:=\hat{A}_x:=\frac{1}{2} (x-1)\frac{\pl}{\pl x},
$$
and consider the partial differential equation (49) with initial
condition (50):
$$
\frac{1}{2} z(\frac{\pl}{\pl z}-1) \Psi(x,z)=\frac{1}{2}
(x-1)\frac{\pl}{\pl x}\Psi(x,z)
$$
$$
\left(4A^2-2A-\nu^2+\frac{1}{4}\right)\Psi(0,z)=z^2\Psi(0,z),~
\mbox{with}~
\Psi(0,z)=1+0(z^{-1}).
$$
{}\hfill (Bessel equation)
\newline The solution to this problem is given
by\footnote{$\varepsilon=i\sqrt{\frac{\pi}{2}} e^{i\pi\nu/2},
-\frac{1}{2}<\nu<\frac{1}{2}$.}
$$
\Psi(0,z)=B(z)=\varepsilon \sqrt{z}
H_{\nu}(iz)=\frac{e^z2^{\nu+\frac{1}{2}}}{\Gamma
(-\nu+\frac{1}{2})} \int_1^{\infty} \frac{z^{-\nu+\frac{1}{2}}
e^{-uz}}{(u^2-1)^{\nu+\frac{1}{2}}} \, du=1+0\left(\frac{1}{z}\right)
$$
and
\be
\Psi(x,z)=e^{xz} B((1-x)z).
\ee
The latter is an eigenfunction for the second order problem
$$
\left(D^2-\frac{\nu^2-\frac{1}{4}}{(x-1)^2}\right)
\Psi(x,z) = z^2 \Psi(x,z).
$$

Choosing in (52), $b_+=e^{-\frac{i\pi\nu}{2}}/2\sqrt{\pi}$,
$b_-=i\bar b_+$, and integrating from 1 to $x=i+1$ leads to the
Bessel kernel mentioned in (7).  If $E$ is a semi-infinite interval,
the first equation of (53) reduces to the Painlev\'e V equation for
$(-A\pl/\pl A) \log \det(I-K^E)$, which was first done
in\cite{TW1}.  Specializing to $\nu=\pm \frac{1}{2}$ leads to the sine
kernels, mentioned in (7). Finally, the $\tau$-function associated
with the wave function is a double Laplace-like transform;
see\cite{AMSV}. The details for the Bessel kernel can be found
in\cite{ASV2}.

 \section*{Acknowledgments}

The
support of National Science Foundation \# DMS 95-4-51179,
Nato, FNRS and Francqui Foundation grants is gratefully
acknowledged.

\end{document}